\documentclass[aps,pra,twocolumn,letterpaper, floatfix, superscriptaddress,amsmath,amssymb]{revtex4-2}
\usepackage{amsmath,amssymb, graphicx}
\usepackage{bm,times, color}
\usepackage[utf8]{inputenc}
\usepackage[colorlinks,breaklinks]{hyperref}
\usepackage[svgnames]{xcolor}
\hypersetup{linkcolor=DarkBlue, citecolor=DarkBlue, filecolor=black, urlcolor=DarkBlue}
\usepackage{comment, mathrsfs,slashed}
\usepackage{amsfonts}
\usepackage{bbm}
\usepackage[normalem]{ulem}
\def \p{\partial}
\def \dag{\dagger}
\def \mb{\mathbf}

\def \lan{\langle}
\def \ran{\rangle}

\def \ga{\gamma}

    \makeatletter
\def\@fnsymbol#1{\ensuremath{\ifcase#1\or \dagger\or \ddagger\or
   \mathsection\or \mathparagraph\or \|\or **\or \dagger\dagger
   \or \ddagger\ddagger \else\@ctrerr\fi}}
    \makeatother

\begin{document}
%\preprint{EFI-xx}
\title{Gauge field fluctuation corrected QED${}_3$ effective action by fermionic particle-vortex duality}
\author{Wei-Han Hsiao}
\altaffiliation{Ph.D. in Physics, The University of Chicago}
\noaffiliation{}

%\affiliation{Independent Researcher, Chicago, IL, USA}
%\author{Yu-Ping Lin}
%\affiliation{Department of Physics, University of Colorado, Boulder, Colorado 80309, USA}
\date{\today}

\begin{abstract}
We present a non-perturbative framework for incorporating gauge field fluctuations into effective actions of QED${}_3$ in the infrared using fermionic particle-vortex duality. This approach is demonstrated through the applications to models containing $N$ species of 2-component Dirac fermions in solvable and interpretable electromagnetic backgrounds, focusing on $N=1$ or $2$. For the $N=1$ model, we establish a correspondence between fermion Casimir energy at finite density and the magnetic Euler-Heisenberg Lagrangian, and further evaluate the corrections to their amplitudes. This predicts amplification of charge susceptibility and reduction of magnetic permeability. We additionally provide physical interpretations for each component of our calculation and offer alternative derivations based on energy density measurements in different characteristic lengths. For $N=2$, we show that magnetic catalysis is erased in a U(1)$\times$U(1) QED${}_3$, indicating no breakdown of chiral symmetry. Reasoning is offered based on the properties of the lowest Landau level wave functions.
\end{abstract}

\maketitle

\section{Introduction}
The method of effective action has been a powerful technique since the early stages of field theories. A particular branch of its applications concerns quantum electrodynamics (QED). One of the seminal examples is the Euler-Heisenberg Lagrangian \cite{1936ZPhy...98..714H}, which initially calculated effective action as a functional of a prescribed constant electromagnetic field by solving the electron degrees of freedom within the framework of (3+1)-dimensional QED. It predicts implications such as dynamics of photon-photon interaction and non-perturbative pair production in vacuum under a strong electric field \cite{PhysRev.82.664}. This calculation was later generalized to scalar QED, finite density \cite{PhysRevD.42.2881, PhysRevD.51.2026} , finite temperature \cite{PhysRevD.19.2385, PhysRevD.60.067703}, smaller spacetime dimensions \cite{PhysRevD.51.R2513, PhysRevD.55.6218}, and more realistic electromagnetic profiles \cite{PhysRevD.2.1191, PhysRevD.58.105022}.

The conventional framework of this type of effective action consists of dynamical matter fields and an electromagnetic field. The latter is considered classical. The path integral only sums over the quantum fluctuations from the matter by inverting the operator equation of motion. As its essence relies upon the solution to operator-valued differential equations, exact results are only feasible for a few profiles of electromagnetic fields, the most renowned ones being configurations of constant electric and magnetic fields. By construction, this framework does not incorporate the quantum mechanical effect of the electromagnetic field. It is naturally tempting to look for an improved machinery to accomplish this goal. As a plausible step, we can consider a dynamical electromagnetic field $F_{\mu\nu}$ with a mean value $\lan F_{\mu\nu}\ran$ and ask what the analogous effective potential is, as a functional of $\lan F_{\mu\nu}\ran$, after integrating out $\delta F_{\mu\nu} = F_{\mu\nu}-\lan F_{\mu\nu}\ran$. This is reminiscent of the one-particle irreducible (1PI) effective potentials of the scalar bosons \cite{PhysRev.127.965, PhysRevD.7.1888}, and the background field method used for QCD beta function evaluation if we proceed with perturbation methodology \cite{weinberg_1996, Peskin:257493}.

This work shall address this task non-perturbatively in (2+1) dimensions for some spinor QEDs via fermionic particle-vortex dualities \cite{PhysRevX.5.031027, PhysRevX.5.041031, PhysRevB.93.245151, PhysRevX.6.031043, SEIBERG2016395}. Within our scope, duality refers to two Lagrangians, $\mathscr L$ and $\widetilde{\mathscr L}$, leading to the same partition function. In (2+1) dimensions, such pairs of Lagrangians can be derived by relativistic flux attachment \cite{goldman2020duality}, or by assuming a master duality and performing SL(2, $\mathbb Z$) transformations on the U(1) gauges fields in the master Lagrangians \cite{witten2003sl, PhysRevX.6.031043, SEIBERG2016395}. In particular, we shall concentrate on ones of the form:
\begin{align}\label{schematicDuality}
& \mathscr L[\{\psi_I\}, A_{\mu}] \leftrightarrow  \widetilde{\mathscr L}[\{\chi_I\}, \{a_{I,\mu}\}, A_{\mu}] \\ \notag
= &   \widetilde{\mathscr L}_1[\{\chi_I\}, \{a_{I,\mu}\}] + \widetilde{\mathscr L}_2[\{a_{I,\mu}\}, A_{\mu}],
\end{align}
where $\psi_I$ and $\chi_I$ are 2-component Dirac fields, $A_{\mu}$ is the background electromagnetic field, and $a_{I,\mu}$ is the emergent gauge field attached to $\chi_I$, with $I$ ranging from $1$ to $N$. Physically, $\chi_I$ represents the vortex of $\psi_I$. By studying the responses to $\delta A_{\mu}$ and the equations of motion of $a_{I,\mu}$, we may determine the mean-field values of $\lan a_{I, \mu}\ran$ by controlling the background $A_{\mu}$. Reading off from the decomposition in Eq.~\eqref{schematicDuality}, $\widetilde{\mathscr L}_1$ merely consists of matter fields $\{ \chi_I\}$ and dynamical gauge fields $\{a_{I, \mu}\}$. Its path integral can be evaluated by separating the contribution of $\widetilde{\mathscr L}_2[\{a_{I,\mu}\}, A_{\mu}]$ from $\mathscr L[\{\psi_I\}, A_{\mu}]$, where the latter shall be computed with more conventional methods. 

We will explicitly execute this recipe in the bulk of this work for concrete dualities with $N = 1, 2$ (referencing Eqs.\eqref{N=1Duality}, and~\eqref{N=2Duality}) in the limit of the infrared (IR) for two specific configurations of $A_{\mu}$, a constant chemical potential and a constant magnetic field, at zero temperature. The separation of $\widetilde {\mathscr L}_2$ is carried out by a Legendre transform over the effective action from $\mathscr L[\{\psi_I\}, A_{\mu}]$. As a result, we will be able to compute the effective actions attributed to $\widetilde{\mathscr L}_1$ in the dual mean-value backgrounds, a constant magnetic field and a constant chemical potential respectively. For $N=1$, the results imply the amplification of Casimir energy at finite density and the magnetic Euler-Heisenberg Lagrangian. These amplifications are both of order 1. Equivalently, these results indicate the amplification of charge susceptibility given the same value of chemical potential. Moreover, we endow the Legendre transform with the physical interpretation of offsetting the local energy density by the amount of chemical potential going from the particle description to the vortex one and vice versa. This picture inspires an alternative derivation for the quantum effective action based on a scaling argument. The dual effective action is the original one offset by the chemical potential and measured in the characteristic scale of the dual description. The characteristic length scale is the inter-particle distance at finite density and the magnetic length in a magnetic field. For $N=2$, we employ this approach to the study of spontaneous breakdown of chiral symmetry in a QED${}_3$ with two U(1) gauge fields, coupling to the sum and the difference of fermion charges respectively. We find that fermion mass condensate produced by a constant magnetic field in the free theory is erased by quantum fluctuations from two U(1) gauge fields. An investigation upon the properties in the lowest Landau level is conducted to understand how symmetry might be enforced. 

In addition to the quantitative predictions, this machinery is intriguing in its own right. Although the concept of the Legendre transform is by no means revolutionary and, in fact, was introduced in the early stages of the background field method, our duality approach is novel in that the functional determinant can be evaluated exactly in the dual theory thanks to the limit of IR, yielding physically interpretable results. Our machinery serves as an attempt to improve the effective action approach which is orthogonal to others such as derivative expansions and higher fermion loops, and aims at capturing the quantum effects of gauge fields. It also opens up another quantitative applications of (2+1) dimensional dualities besides response functions, such as electric and thermal transport properties under various deformations \cite{PhysRevB.96.075127, PhysRevB.96.245140, PhysRevB.100.235150, PhysRevD.104.125006, PhysRevB.108.235142}. 

The rest of the paper shall be organized as follows. In Sec.\ref{framework}, we provide an essential review for the specific dualities and an overview of the questions of our interest. On top of these, we lay out the derivation of the effective actions and some key formulae. Sec.\ref{result} is devoted to employing the machinery for explicit computations as well as the physical interpretations of the results. Some recap and open directions are presented in Sec.\ref{conclusion}. To be self-contained, the appendix elaborates the computational details we referenced in this work.

\section{Effective action from particle vortex duality}\label{framework}
\subsection{$N=1$ duality}
Let us first consider 2-component Dirac spinors in (2+1) dimensions. The fermionic particle-vortex duality states that a free Dirac field $\psi$ is dual to a QED${}_3$ of another Dirac field $\chi$ via a mixed Chern-Simons term: 
\begin{align}\label{N=1Duality}
 i \bar{\psi}\slashed{D}_A\psi \leftrightarrow i\bar{\chi}\slashed{D}_a \chi + \frac{1}{4\pi} \epsilon^{\mu\nu\lambda}a_{\mu}\p_{\nu}A_{\lambda},
\end{align}
where $(D_A)_{\mu}= \p_{\mu} - iA_{\mu}$ refers to the covariant derivative associated with the gauge field $A_{\mu}$. Both $A_{\mu}$ and $a_{\mu}$ are U(1) gauge fields. $A_{\mu}$ is a classical background field, whereas $a_{\mu}$ is emergent and dynamical. The arrow ``$\leftrightarrow$'' indicates that the partition functions from both sides of \eqref{N=1Duality} are equivalent as functionals of $A_{\mu}$:
\begin{align}
 e^{i\mathscr W[A] } &= \int \mathscr D\bar{\psi}\mathscr D\psi e^{i\int d^3x\, ( i \bar{\psi}\slashed{D}_A\psi )} \notag\\
& =  \int \mathscr Da\mathscr D\bar{\chi}\mathscr D\chi\, e^{i\int d^3x\,( i\bar{\chi}\slashed{D}_a \chi + \frac{1}{4\pi} \epsilon^{\mu\nu\lambda}a_{\mu}\p_{\nu}A_{\lambda})}.
\end{align}
The statement~\eqref{N=1Duality} is a fermionic analogue of the 3D-XY model and Abelian Higgs duality \cite{PESKIN1978122, PhysRevLett.47.1556} and was proposed in the studies of the metallic state in the fractional quantum Hall effect and the surface states of topological insulators \cite{PhysRevX.5.031027, PhysRevX.5.041031, PhysRevB.93.245151}. Note that it is well-known in (2+1) dimensions a single massless Dirac cone cannot properly define the partition function without breaking either gauge invariance or parity \cite{PhysRevD.29.2366, RevModPhys.88.035001}. This inconsistency can be remedied by including a half-level Chern-Simons term $\frac{1}{8\pi}\epsilon^{\mu\nu\lambda}A_{\mu}\p_{\nu}A_{\lambda}$ to the Lagrangian in \eqref{N=1Duality}. In the following of the work, this contact term will be included implicitly into the convention of the path integral because it will not concern the main arguments and the results, which shall all be parity-even quantities.

A set of operator correspondences can be derived by varying Eq.~\eqref{N=1Duality} with respect to the gauge fields $A_{\mu}$ and $a_{\mu}$. On one hand, varying the probe $A_{\mu}$ on the both sides of the duality identifies the particle current of $\psi$ with the dual field strength tensor of $a_{\mu}$:
\begin{subequations}
\begin{align}\label{mapping1}
\bar{\psi}\ga^{\mu}\psi = \frac{1}{4\pi}\epsilon^{\mu\nu\lambda}\p_{\nu}a_{\lambda}.
\end{align}
On the other hand, varying the dual Lagrangian with respect to $a_{\mu}$ leads to the equation of motion that imposes a constraint on particle current of $\chi$ and the dual field strength of $A_{\mu}$:
\begin{align}\label{mapping2}
\bar{\chi}\ga^{\mu}\chi = -\frac{1}{4\pi}\epsilon^{\mu\nu\lambda} \p_{\mu}A_{\lambda}.
\end{align}
\end{subequations}
The zeroth components of Eqs.~\eqref{mapping1} and~\eqref{mapping2} imply the following relations \footnote{We adopt the convention $B = (\nabla\times\mb A)$ and $A_{\mu} = (A_0, -\mb A)$.}:
\begin{subequations}
\begin{align}
\label{mapping3}& n_{\psi} = -\frac{b}{4\pi},\, b = \partial_xa^2 - \partial_y a^1\\
\label{mapping4}& n_{\chi} = \frac{B}{4\pi},\, B = \partial_xA^2 - \partial_yA^1.
\end{align}
\end{subequations}
These are the typical signatures of particle-vortex dualities, where the charge density on one side of a duality becomes a magnetic field on the other. Relatedly, the spatial components of Eqs~\eqref{mapping1} and~\eqref{mapping2} entail that the current density maps to rotated electric field under duality. In a gedankenexperiment, when we turn on an external magnetic field $B$, a charge density profile of $\chi$ would form accordingly, implying a finite chemical potential $a_0$. Similarly, when we turn on a chemical potential $A_0$ to develop a charge density profile of $\psi$, it in turn creates a mean magnetic field $b$ in the dual description.

Eq.~\eqref{N=1Duality} and the correspondences between chemical potential and magnetic field constitute the main motivation of this work, as the free Dirac cone is exactly solvable under certain background field profiles. Particularly, given the background field configuration with constant $B$ and $A_0$, the effective action $\mathscr W[A_0, B]$ is formally 
\begin{widetext}
\begin{align}\label{effSDuality}
&\mathscr W[A_0, B]=  - i\log \int \mathscr D\bar{\psi}\mathscr D\psi\, \exp\left({i\int d^3x\,  i \bar{\psi}\slashed{D}_A\psi }\right) \notag\\
=&  -i\log \int \mathscr Da\mathscr D\bar{\chi}\mathscr D\chi\, \exp\left({i\int d^3x\, i\bar{\chi}\slashed{D}_a\chi - \frac{i\mathscr V_3}{4\pi}\bar{a}_0 B - \frac{i\mathscr V_3}{4\pi}\bar bA_0}\right)\notag\\
= & -i\lim_{g\to \infty}\log \int \mathscr Da\mathscr D\bar{\chi}\mathscr D\chi\, \exp\left({i\int d^3x\, i\bar{\chi}\slashed{D}_a\chi - \frac{i\mathscr V_3}{4\pi}\bar{a}_0 B - \frac{i\mathscr V_3}{4\pi}\bar bA_0 - \frac{i}{4g^2}\int (\p_{\mu}a_{\nu}-\p_{\nu}a_{\mu})^2}\right),
\end{align}
where the barred quantities are spacetime average over the spacetime volume $\mathscr V_3$: $
 \bar a_0 = {\mathscr V_3}^{-1}\int d^3x\, a_0$ and $ \bar b = {\mathscr V_3}^{-1}\int d^3x\, b$.
In the last line of Eq.~\eqref{effSDuality}, we reintroduce the Maxwell dynamics for the emergent gauge field $a_{\mu}$ and interpret the dual description as the infrared limit $g\to\infty$ of the QED${}_3$. As explained, applied constant $B$ and $A_0$ would induce corresponding average values of $a_0$ and $b$ in the dual theory. Let us now consolidate the $\bar{\chi}\slashed{D}_a\chi$ and the Maxwell term $(\p_{\mu}a_{\nu}-\p_{\nu}a_{\mu})^2$ into $\mathscr L_{\rm QED_3}$. Eq.~\eqref{effSDuality} formally represents QED${}_3$ coupled to two external sources $J_1 = B$ and $J_2 = A_0$. Varying $\mathscr W$ with respect to $J_1$ and $J_2$ leads to the associated field expectation values of $a_0$ and $b$:
\begin{subequations}
\begin{align}
 \mathscr V_3^{-1}\frac{\p\mathscr W}{\p B} &= -\frac{1}{4\pi}e^{-i\mathscr W}\int \mathscr Da\mathscr D\bar{\chi}\mathscr D\chi\, \bar a_0 \exp\left({i\int d^3x\, \mathscr L_{\rm QED_3}[\chi, a_{\mu}] - \frac{i\mathscr V_3}{4\pi}\bar{a}_0 B - \frac{i\mathscr V_3}{4\pi}\bar bA_0 }\right) = -\frac{1}{4\pi}\lan \bar a_0\ran := -\frac{1}{4\pi}a_c\\
 \mathscr V_3^{-1}\frac{\p\mathscr W}{\p A_0} & = -\frac{1}{4\pi}\lan \bar b\ran := -\frac{1}{4\pi}b_c. 
\end{align}
\end{subequations}
\end{widetext}
Following the standard argument for quantum effective potential \cite{Peskin:1995ev}, after integrating $\chi$ and $a_{\mu}$, the 1PI effective action $\mathscr V_3\Gamma$ of $\mathscr L_{\rm QED_3}$, as a function of $a_c$ and $b_c$, should be obtained via a Legendre transform of $\mathscr W$ with respect to $B$ and $A_0$, and is expected to assume the form:
\begin{align}\label{Legendre1}
\mathscr V_3^{-1}{\mathscr W[A_0, B]}= \Gamma[a_c, b_c] - \frac{1}{4\pi}a_cB - \frac{1}{4\pi}b_cA_0.
\end{align}
By matching the path integral~\eqref{effSDuality}, it can be inferred that $\Gamma[a_c, b_c]$ physically corresponds the IR effective Lagrangian density of the QED${}_3$
\begin{align*}
\lim_{g\to\infty}\left[ i \bar{\chi}\slashed{D}_a\chi - \frac{1}{4g^2}(\p_{\mu}a_{\nu}-\p_{\nu}a_{\mu})^2\right],
\end{align*} 
evaluated in the presence of a classical configuration of $a_c$ and $b_c$, incorporating fluctuations from both $\chi$ and $a_{\mu}$. It is worth noting that the background-field terms in Eq.~\eqref{Legendre1} neutralizes the charge of $a_{\mu}$ when $\lan\bar a_0\ran = a_c$ is finite, because its full current is $J^{\mu}_a = \bar{\chi}\ga^{\mu}\chi + \frac{1}{4\pi}\epsilon^{\mu\nu\lambda}\p_{\nu}A_{\lambda}$ for finite $g^2$. Consequently, in the limit $g\to \infty$, the Coulomb term $\sim g^2J_{\mu}\frac{1}{\p^2}J^{\mu}$ is rendered by the saddle point solution~\eqref{mapping2}. The Legendre transform is performed on the entirety of the right-hand side of Eq.\eqref{Legendre1}.

Should we continue with the conventional route toward the quantum effective action, we would expand fields in $\mathscr L_{\rm QED_3}$ about $a_c$ and $b_c$ to quadratic order, integrate the fermion and the gauge field in the Gaussian approximation, and solve the Legendre transform perturbatively. In our case, particle-vortex duality~\eqref{N=1Duality} offers an alternative and more precise path, as $\mathscr W$ can be evaluated using $i\bar{\psi}\slashed D_{A}\psi$ by integrating $\psi$ exactly. In other words, the effect of gauge field path integral can be computed by Legendre transforming the effective action $\mathscr W$, or equivalently the effective energy density $\mathscr  E[A_0, B] = -\mathscr V_3^{-1}\mathscr W[A_0, B] $. To utilize this formulation, we may first compute $\mathscr E$ or $\mathscr W$ using approaches such as Schwinger proper time \cite{PhysRev.82.664, PhysRevD.42.2881, PhysRevD.51.2026, PhysRevD.55.6218}, and then perform the Legendre transform by solving $B, A_0$ in $a_c, b_c$ using the relations:
\begin{subequations}
\begin{align}
& \frac{\p\mathscr  E}{\p B} = \frac{a_{c}}{4\pi},\\
& \frac{\p\mathscr  E}{\p A_0} = \frac{b_{c}}{4\pi}.
\end{align}
\end{subequations}
Plugging the solutions back into~\eqref{Legendre1} yields the target effective action for QED${}_{3}$:
\begin{align}
\label{masterFormulaN=1}\Gamma[a_{c}, b_{c}] = &- \mathscr E[A_0(a_{c}, b_{c}), B(b_{c}, a_{c})] \notag\\
& +  \frac{a_{c}}{4\pi}B(b_{c}, a_{c}) + \frac{b_{c}}{4\pi}A_0(b_{c}, a_{c}).
\end{align}
We shall exploit this correspondence to evaluate the effect of gauge field fluctuation on the resulting quantum effective action.
\subsection{$N=2$ duality}
Similar arguments can be generalized to $N$ species of fermion fields to derive gauge field fluctuation-included effective action. An $N=2$ example can be motivated by a 4-component Dirac Lagrangian in (2+1) dimensions
\begin{align}
\label{N=2free}\mathscr L_{N=2} = i\bar{\Psi}\Gamma^{\mu}(\p_{\mu} - i A_{\mu})\Psi, \Gamma^{\mu} = \begin{pmatrix} \ga^{\mu} & 0 \\ 0 & \bar{\gamma}^{\mu}\end{pmatrix},
\end{align}
where $\ga^{\mu}$ and $\bar{\ga}^{\mu}$ are two inequivalent irreducible representations of 2 by 2 gamma matrices. By writing $\Psi = (\psi_1, \psi_2)^T$, the 4-component Lagrangian can be expressed as:
\begin{align}\label{mainN=2}
\mathscr L_{N=2} = \bar{\psi}_1\ga^{\mu}iD_{\mu}\psi_1 + \bar{\psi}_2\bar{\ga}^{\mu}iD_{\mu}\psi_2.
\end{align}
Note that the ``bar'' for $\psi_1$ and $\psi_2$ fields are defined by different gamma matrices $\bar{\psi}_1 = \psi^{\dag}_1\ga^0$ and $\bar{\psi}_2 = \psi^{\dag}\bar{\ga}^0$. The dual theory is derived by dualizing $\psi_1$ and $\psi_2$ separately:
\begin{align}
\label{N=2Duality}\mathscr L_{N=2} \leftrightarrow   \bar{\chi}_1\ga^{\mu}i & D_{\mu}[a_1]\chi_1  + \bar{\chi}_2\bar{\ga}^{\mu}iD_{\mu}[a_2]\chi_2\notag\\
& + \frac{1}{4\pi}\epsilon^{\mu\nu\lambda}(a_1 + a_2)_{\mu}\p_{\nu}A_{\lambda}.
\end{align}
It is crucial that there are two emergent gauge fields $a_1$ and $a_2$. The SU(2) structure from the $\Psi$ perspective is not apparent in terms of the dual Dirac fields $\chi_1, \chi_2$, for which only U(1)$\times$U(1) manifests. To establish the operator mapping, let us use the following variables $a = (a_1 + a_2)/ 2$, $\alpha = (a_1 - a_2)/2$, and $X = (\chi_1, \chi_2)^T$ to rephrase the dual theory:
\begin{align}
\label{DualN=2}\widetilde{\mathscr L}_{N=2} = i\bar{X}\Gamma^{\mu}D_{\mu}[a]X & + \alpha_{\mu}(\bar{\chi}_1\ga^{\mu}\chi_1 - \bar{\chi}_2\bar{\ga}^{\mu}\chi_2)\notag\\
& + \frac{1}{2\pi}\epsilon^{\mu\nu\lambda}a_{\mu}\p_{\nu}A_{\lambda}.
\end{align}
By the same token, the operator correspondences can be derived by varying both Eqs.~\eqref{mainN=2} and~\eqref{DualN=2} with respect to $A_{\mu}$ and imposing the equation of motion of $a_{\mu}$ on Eq.~\eqref{DualN=2}:
\begin{subequations}
\begin{align}
\label{mapping5}& \bar{\Psi}\Gamma^{\mu}\Psi = \frac{1}{2\pi}\epsilon^{\mu\nu\lambda}\p_{\nu}a_{\lambda}\\
\label{mapping6}& \bar X\Gamma^{\mu}X = -\frac{1}{2\pi}\epsilon^{\mu\nu\lambda}\p_{\nu}A_{\lambda}.
\end{align}
The second U(1) field $\alpha_{\mu}$ demands the other constraint:
\begin{align}
\label{mapping7} \bar{\chi}_1\ga^{\mu}\chi_1 = \bar{\chi}_2\bar{\ga}^{\mu}\chi_2.
\end{align}
\end{subequations}
Similar to Eqs.~\eqref{mapping1} and~\eqref{mapping2}, the zeroth components of Eqs.~\eqref{mapping5} and~\eqref{mapping6} entail that a finite mean magnetic field on one side of duality corresponds to a finite total particle density on the other side, which in turn induces a mean chemical potential. As a consequence, for this particular U(1)$\times$U(1) QED${}_3$model we can write down its effective Lagrangian as a function of classical values of chemical potential $a_{c}$ and magnetic field $b_{c}$ in terms of the Legendre transformation of the effective potential of a 4-component Dirac field in a classical electromagnetic background:
\begin{align}
& \Gamma[a_{c}, b_{c}] = -\mathscr E[A_0, B] + \frac{a_{c}}{2\pi}B + \frac{b_{c}}{2\pi}A_0
\end{align}
along with the following identities 
\begin{subequations}
\begin{align}
& \mathscr E= \frac{i}{\mathscr V_3}\ln\int \mathscr D\bar{\Psi}\mathscr D\Psi \exp\left( i \int d^3x\, \mathscr L_{N=2}[\Psi, A]\right)\\
\label{N=2BIna} & \frac{\p \mathscr E}{\p B} = \frac{1}{2\pi}a_{c},\\
\label{N=2AInb}&  \frac{\p \mathscr E}{\p A_0} = \frac{1}{2\pi}b_{c} .
\end{align}
\end{subequations}
The $N=2$ theory is peculiar compared to its $N=1$ counterpart in few ways. We highlight a couple of our main interest as follows. Firstly, within the 4-component spinor framework, it is possible to construct a chirality operator $\ga_5$ that anti-commutes with all other $\Gamma^{\mu}$. This $\ga_5$ introduces the concepts of left-handed and right-handed fermions as its eigenstates. Secondly, we could construct a parity-even mass $\bar{\Psi}\Psi$ as opposed to a single 2-component Dirac field. Nevertheless, this mass term breaks chiral symmetry. To illustrate, one explicit representation \cite{PhysRevD.52.4718} of $\Gamma^{\mu}$, $\ga_5$ and the parity transformation $\mathsf P$ using the standard Pauli matrices $(\sigma_1, \sigma_2, \sigma_3)$ and 2 dimensional identity matrix $I_2$ goes as 
\begin{subequations}
\begin{align*}
& \Gamma^{\mu} =\left( \begin{pmatrix} \sigma_3 & 0 \\ 0  & -\sigma_3\end{pmatrix}, \begin{pmatrix} i \sigma_1 & 0 \\ 0 & -i\sigma_1\end{pmatrix}, \begin{pmatrix} i \sigma_2 & 0 \\ 0 & -i\sigma_2\end{pmatrix}\right),\\
& \ga_5 = i \begin{pmatrix} 0 & I_2\\ - I_2 & 0 \end{pmatrix},\\
& \mathsf P: \Psi(x^0, x^1, x^2)\to \begin{pmatrix} 0 & -i\sigma_1 \\ i\sigma_1 & 0 \end{pmatrix}\Psi(x^0, -x^1, x^2).
\end{align*}
\end{subequations}
This representation can be regarded as the (2+1) dimensional analogue of the Dirac representation, where the mass matrix $\Gamma^0$ is diagonal and $\ga_5$ is off-diagonal. In terms of the spinors $\psi_I$, the chiral current mixes different species $J^{\mu}_5 = \bar{\Psi}\Gamma^{\mu}\ga_5\Psi = i (\bar{\psi}_1\ga^{\mu}\psi_2 - \bar{\psi}_2\bar{\ga}^{\mu}\psi_1).$ The same property applies to the $X$ field. It can be confirmed by direct calculation that $\bar{\Psi}\Psi = \Psi^{\dag}\Gamma^0\Psi$ is even under $\mathsf P$, whilst under $\Psi \to e^{i\beta\ga_5}\Psi$
\begin{align*}
\delta (\bar{\Psi}\Psi)  = 2i\beta\bar{\Psi}\ga_5\Psi = 2i\beta (\bar{\psi}_1\psi_2- \bar{\psi}_2\psi_1).
\end{align*}
 These two properties compose a series of questions concerning the spontaneous breakdown of chiral symmetry \cite{PhysRevD.33.3704}. That is, given a vanishing fermion mass, under what circumstance would a finite expectation value $\lan \bar{\Psi}\Psi\ran$ persist? One well-known example is the magnetic catalysis \cite{PhysRevLett.73.3499, PhysRevD.52.4718, DITTRICH1997182} for a 4-component Dirac fermion in (2+1) dimensions in a constant magnetic field $B$. Fermion mass bilinear $|\lan\bar{\Psi}\Psi\ran|$ converges to a finite value $|B|/(2\pi)$ in the massless limit.  

The $N=2$ duality provides a device to explore the phenomena of spontaneous symmetry breakdown such as the magnetic catalysis in interacting theories. To investigate the fermion condensate in this framework, we first remind ourselves that the vacuum expectation value of the mass bilinear can be obtained by taking the derivative with respect to mass over the effective action of the target massive Dirac field:
\begin{align}
 & e^{i\mathscr W} = \int\mathscr D\bar{\Psi}\mathscr D\Psi\, \exp\left( i \int d^3x\, \bar{\Psi}(i\slashed D - M)\Psi\right)\notag\\
 \Rightarrow & \frac{\p \mathscr W}{\p M} = - \int d^3x\lan \bar{\Psi}\Psi\ran.
\end{align}
Hence the goal is to derive the quantum effective action consisting of the mass parameter. For Eq.~\eqref{N=2free}, computing the massive fermion determinant is fairly straightforward. In order to take the mass derivative on the dual effective action, we require the mass operator correspondence in the dual theory. In the process of Legendre transform, the ordinary mass parameter has to be solved in terms of the dual mass. To that end, let us elucidate the derivation of the mass operator correspondence by deforming the $N=1$ duality~\eqref{N=1Duality}. The $N=2$ theory is simply two copies of it. 

Suppose we deform the left-hand side of Eq.~\eqref{N=1Duality} with a mass term $m_{\psi}\bar{\psi}\psi$. At low energies, the leading quadratic effective action is $\frac{\mathrm{sgn}(m_{\psi})}{8\pi}\epsilon^{\mu\nu\lambda}A_{\mu}\p_{\nu}A_{\lambda}$. In order to obtain the same effective Lagrangian from the dual description, we would need to deform the right-hand side of Eq.~\eqref{N=1Duality} with another mass term $m_{\chi}\bar{\chi}\chi$, where $\mathrm{sgn}(m_{\chi}) = -\mathrm{sgn}(m_{\psi})$. By the same token we can conclude the mass operator on one side of the duality maps also to the mass operator on the other side, yet with a different sign
\begin{align}
M_{\Psi}\bar{\Psi}\Psi \leftrightarrow M_X\bar XX, \mathrm{sgn}(M_{\Psi}) = -\mathrm{sgn}(M_X).
\end{align}
Putting them all together, we are able to evaluate the fermion condensate in the U(1)$\times$U(1) QED${}_3$ up to a proportionality constant through the formula 
\begin{align}\label{condensateEq}
 & \lan \bar XX\ran = -\lim_{M_{X}\to 0}\frac{\p \Gamma[a_c, b_c]}{\p M_{X}} \notag\\
\sim & \lim_{M_{\Psi}\to 0}\frac{\p}{\p M_{\Psi}} \left[-\mathscr E + \frac{1}{2\pi}a_c B + \frac{1}{2\pi} b_cA_0\right].
\end{align}
We shall apply it again to the examples of constant chemical potential and constant magnetic field.
\section{Explicit examples and results}\label{result}
In what follows we will zoom into specific examples and compute physical quantities using the tools developed in the preceding section. During the course of the derivation, we presented chemical potential and magnetic field simultaneously. Nevertheless, in the rest of the work we shall render the results more physically interpretable by turning on one of them at a time.
\subsection{$N=1$ duality}
\subsubsection{$A_0\neq 0, B=0$}
This background field produces a finite density $n_{\psi}$ for $\psi$ and a finite background magnetic field $\lan \p_1a^2 - \p_2a^1\ran = b$ for $\chi$. $\psi$ theory is free and forms a Fermi disk. Meanwhile, the mean-field ground state of $\chi$ is the charge-neutral lowest Landau level, coupling to the fluctuating $a_{\mu}$.

One way to compute the effective energy for $\psi$ is by identifying the ground state particle density with the derivative of energy density with respect to the chemical potential,
\begin{align}
\frac{\p \mathscr E}{\p A_0} = -n_{\psi}.
\end{align}
The particle density for a massless free Dirac fermion is 
\begin{align}\label{psiDensity}
n_{\psi} =\mathrm{sgn}(A_0) \frac{p_F^2}{4\pi} = \mathrm{sgn}(A_0)\frac{A_0^2}{4\pi},
\end{align}
implying:
\begin{align*}
\mathscr E = -\int^{A_0}_0 d\nu\, n_{\psi}(\nu) = -\frac{1}{12\pi}\mathrm{sgn}(A_0)A_0^3 = -\frac{|A_0|^3}{12\pi}.
\end{align*}
Negative $A_0$ corresponds to the density of antiparticles. $\mathscr E$ can also be evaluated by computing a functional determinant:
\begin{align*}
 \mathscr E& = \frac{i}{\mathscr V_3}\left( \mathrm{Tr}\ln[i\slashed{\p} + A_0\ga^0] - \mathrm{Tr}\ln i\slashed{\p}\right)\notag\\
 & = \frac{i}{\mathscr V_3} \int^{A_0}_0 d\nu\,  \mathrm{Tr}[G(0; \nu)\ga^0],
\end{align*}
where we applied the trick $\mathrm{Tr}\frac{\p}{\p A_0}\ln[i\slashed{\p} + A_0\ga^0] = \mathrm{Tr}[(i\slashed{\p} + A_0\ga^0)^{-1}\ga^0] = \mathrm{Tr}[G(0; A_0)\ga^0]$ with the Green's function $(i\slashed{\p} + A_0\ga^0)G(x, y) = \delta(x-y)$. We include the details in the appendix for the purpose of comprehensiveness. That approach will also be utilized when we address $N=2$ models.

With either technique we find:
\begin{align}\label{CasimirA0}
\mathscr E=  \frac{-1}{12\pi}|A_0|^3.
\end{align}
This quantity is negative regardless of the sign of the chemical potential. The physical reason is that by subtracting the vacuum value $A_0 = 0$, our computation results in the Casimir energy of the system -- It sums up $(|\mb p| - A_0)$ for each particle inside the Fermi disk $|\mb p| <p_F$. Equivalently, it implies the fermion pressure is positive definite because the energy density at zero temperature equals the negative of pressure. 

Next we solve $A_0$ in terms of $b_{c}$.
\begin{align}
 \label{A0inb} \frac{\p\mathscr E}{\p A_0} & = -\mathrm{sgn}(A_0) \frac{A_0^2}{4\pi} = \frac{b_{c}}{4\pi}\notag\\
\Rightarrow A_0 & = -\mathrm{sgn}(b_{c})\sqrt{|b_{c}|}. 
\end{align}
This results in:
\begin{align}\label{magneticb}
\Gamma = -\mathscr E + \frac{1}{4\pi}b_cA_0= -\frac{1}{6\pi}|b_{c}|^{3/2}.
\end{align}
This effective action is negative definite regardless of the sign of $b_{c}$. This entails that the magnetic energy in the interacting theory is positive definite, and there is no vacuum decay led by fluctuation. Besides, Eqs~\eqref{psiDensity} and~\eqref{A0inb} are consistent with the operator mapping~\eqref{mapping3}.
\subsubsection{$A_0 = 0, B\neq 0$}
This background situates $\psi$ at the charge-neutral point on the lowest Landau level. For $\chi$, it establishes a mean-field Fermi disk interacting with $a_{\mu}$.

The effective action $\mathscr W[B]$ is described by the renowned Euler-Heisenberg effective Lagrangian \cite{1936ZPhy...98..714H, PhysRev.82.664, PhysRevD.51.R2513, PhysRevD.55.6218}. In the massless limit, the proper-time integral can be evaluated exactly, resulting in 
\begin{align}\label{magneticB}
-\mathscr E =  -\frac{\zeta(3/2)}{4\sqrt{2}\pi^2}|B|^{3/2} = -k |B|^{3/2}.
\end{align}
It is noteworthy that there is a negative sign in the Minkowski signature to ensure $\mathscr E > 0$, and thus the ground state is stable in a constant $B$ \cite{PhysRevD.51.R2513}. For completeness, we also provide a detailed calculation in full Minkowski signature in the appendix.

Solving $B$ in terms of $a_{c}$, 
\begin{align}
\frac{\p\mathscr E}{\p B}= \mathrm{sgn}(B)\frac{3}{2}k |B|^{1/2} = \frac{1}{4\pi}a_{c},
\end{align}
and plugging it back to the full effective action lead to
\begin{align}\label{Casimira0}
\Gamma =  -\mathscr E + \frac{1}{4\pi}a_{c}B = \frac{4}{27k^2}\left( \frac{|a_{c}|}{4\pi}\right)^3.
\end{align}
This quantity is positive definite. In other word, the Casimir energy of $\chi$ at finite density with respect to vacuum is again negative. Following the same recipe, we could show that the sign of the Casimir energy depends upon the sign of the magnetic energy. Comparing Eqs.~\eqref{CasimirA0},~\eqref{magneticb},~\eqref{magneticB}, and~\eqref{Casimira0}, the positive definiteness of the magnetic energy in a constant magnetic background is the dual statement of the negative definiteness of the fermion Casimir energy at finite density relative to the ground state.
As another sanity check, the operator mapping~\eqref{mapping4} is fulfilled. 
\begin{align}
n_{\chi} = \frac{\p\Gamma}{\p a_{c}} = \mathrm{sgn}(a_{c})\frac{4}{9k^2}\left( \frac{a_{c}}{4\pi}\right)^2 \frac{1}{4\pi} = \frac{B}{4\pi}.
\end{align}
\subsubsection{Discussion}
In the preceding part of this section, we have computed the negative Casimir energy of a finite-density Dirac fermion~\eqref{CasimirDual} and the negative magnetic energy of a zero-density Dirac fermion in a mean magnetic field~\eqref{MagneticDual}. Both incorporate the effect of a fluctuating U(1) gauge field:
\begin{subequations}
\begin{align}
\label{CasimirDual}& \Gamma(a_{c})  = \frac{2\pi}{27\zeta^2(3/2)}|a_{c}|^3\\
\label{MagneticDual}& \Gamma(b_{c}) = -\frac{1}{6\pi} |b_{c}|^{3/2}.
\end{align}
\end{subequations}
The signs of these quantities were given special care to justify the correct signs of their free theory counterparts~\eqref{CasimirA0} and~\eqref{magneticB}. In fact, we could relate them with a simple picture by recalling what these energy densities really refer to. In the example $A_0\neq 0, B=0$, $\mathscr E[A_0]$ represents an accumulation of quasiparticle energy relative to the chemical potential up to the Fermi level. The negativeness is due to the displacement from the chemical potential and quasiparticle energy itself is positive-definite. It manifests using the decomposition:
\begin{align*}
\mathscr E[A_0] &= \int^{A_0}_0 (\epsilon - A_0)\frac{\p n_{\psi}}{\p\epsilon}d\epsilon = \int^{A_0}_0 \epsilon \frac{\p n_{\psi}}{\p\epsilon}d\epsilon - A_0 n_{\psi}.
\end{align*}
Together with the Eqs.~\eqref{masterFormulaN=1} and~\eqref{mapping3}, its dual~\eqref{MagneticDual} reads
\begin{align*}
& \Gamma[b_{c}] = -\mathscr E[A_0] + \frac{1}{4\pi}A_0b_{c}\notag\\
 =& -\int^{A_0}_0\epsilon\frac{\p n_{\psi}}{\p\epsilon} d\epsilon + A_0 \left(n_{\psi} + \frac{1}{4\pi}b_{c}\right) =  -\int^{A_0}_0 \epsilon\frac{\p n_{\psi}}{\p\epsilon} d\epsilon.
\end{align*}
It indicates the magnetic energy corresponds to precisely the sum of total quasiparticle excitation energy. Therefore, the magnetic energy $-\Gamma[b_{c}]$ is positive definite. This perspective motivates another interpretation of the result. For a massless free 2-component Dirac fermion, its energy density in terms of the particle density is:
\begin{align}\label{freeTheoryEnergy}
\int \epsilon \frac{\p n_{\psi}}{\p\epsilon} d\epsilon= \frac{p_{\psi, F}^3}{6\pi} = \frac{(4\pi)^{3/2}}{6\pi}n^{3/2}_{\psi}. 
\end{align}
The crucial length scale $r_{\psi}$, the inter-particle distance, is given by $r_{\psi} = n^{-1/2}_{\psi}$. Thus Eq.~\eqref{freeTheoryEnergy} states that there is an amount of energy $\frac{(4\pi)^{3/2}}{6\pi}r^{-1}_{\psi}$ within the area $r^{2}_{\psi}$. In the dual description the natural length scale is the magnetic length $\ell_b = \sqrt{4\pi} \times r_{\psi} = \sqrt{b_{c}}$. The statement is rewritten as there is an amount of energy $\frac{1}{6\pi}\ell^{-1}_b$ within the area $\ell^{2}_b$, or 
\begin{align}
\frac{1}{\ell^2_b}\frac{1}{6\pi\ell_b} = \frac{1}{6\pi}|b_{c}|^{3/2}.
\end{align}
As for the example $A_0=0, B\neq 0$, we can also interpret $\Gamma[a_{c}]$ using a similar decomposition 
\begin{align}
&\Gamma[a_{c}] = -\int^{a_{c}}_0 (\epsilon - a_{c})\frac{\p n_{\chi}}{\p \epsilon} d\epsilon\notag\\
=& -\int^{a_{c}}_0  \epsilon \frac{\p n_{\chi}}{\p \epsilon} d\epsilon+ a_{c} n_{\chi} = -\mathscr E(B) + \frac{a_{c}}{4\pi}B.
\end{align}
From it, we can infer the quasiparticle density of state ${\p n_{\chi}}/{\p \epsilon}$ corrected by the gauge field fluctuation, and it will be derived momentarily. Here the characteristic length scales are the magnetic length $\ell_B =  B^{-1/2}$ and the inter-particle distance $r_{\chi} = n^{-1/2}_{\chi}$. Again using the above integral and the picture of measuring energy density with different length scales, the fermion Casimir energy at finite density $n_{\chi}$ in the dual description amounts to the subtraction of the chemical potential $a_{c}$ per $r_{\chi}^2$ from the magnetic energy per $\ell_B^2$:
\begin{align*}
\frac{1}{\ell^2_B} k\ell^{-1}_B - \frac{ a_{c}}{r_{\chi}^2} = \frac{1}{\ell_B^2}\left(k\ell_B^{-1} - \frac{a_{c}}{4\pi}\right) = \frac{4}{9k^2}\left( \frac{a_{ c}}{4\pi}\right)^3\left[ \frac{2}{3} - 1\right].
\end{align*}
The Interpretation of characteristic length can also help estimate the effective Fermi velocity in the interacting theory. Using Luttinger theorem\footnote{Technically, in the interacting theory it is only a marginal Fermi liquid.}, $n_{\chi} = \frac{1}{4\pi}p_{\chi, F}^2$ but the dispersion is expected to be corrected $a_{c} = vp_{\chi, F}$. The linear relation arises from the fact that there is only one length scale, $r_{\chi}$, in the dual description. Identifying the density, we see $r_{\chi}^{-1} = \frac{a_{c}}{\sqrt{4\pi} v}$ and the magnetic energy is 
\begin{align}
k\ell_B^{-3} = (4\pi)^{3/2}kr_{\chi}^{-3} = \frac{k}{v^3}a_{c}^3,
\end{align}
which should equal $\frac{4}{9k^2}\left( \frac{a_{c}}{4\pi}\right)^3\left[ \frac{2}{3} \right]$, leading to
\begin{align}
v = 4\pi \times \frac{3}{2}k = 0.8820.
\end{align}
Switching the focus back to the results~\eqref{CasimirDual} and~\eqref{MagneticDual}, the functional dependences are not particularly surprising, as they are constrained by dimensional analysis, given there is only one dimensionful quantity in each scenario. However, they quantify the effects of gauge field fluctuations on the magnitude of the energy densities. For instance, with the same level of background magnetic field, there is an enhancement of the ground state magnetic energy due to the quantum correction from $a_{\mu}$:
\begin{align}
\frac{-\Gamma/{|b_c|^{3/2}} }{\mathscr E /{|B|^{3/2}}}  = \frac{2\sqrt 2\pi}{3\zeta(3/2)} = 1.1338.
\end{align}
Schematically, this also implies the amplification of the pair-production or vacuum decay probability via Schwinger mechanism if a constant electric field is turned on because the Euler-Heisenberg Lagrangian is merely modified with $|B|\to \sqrt{B^2 - E^2}$.

In addition, we can compute the correction to the charge susceptibility, or the density of state, $\chi = \frac{\p n}{\p A_0}$, of the Dirac fermions given the same level of chemical potential:
\begin{align}
\frac{a^{-1}_{c}\chi_{\chi}}{A^{-1}_{0}\chi_{\psi}} = \frac{a^{-1}_{c}\frac{\p n_{\chi}}{\p a_{c}}}{A_0^{-1}\frac{\p n_{\psi}}{\p A_0}}= \frac{8\pi^2}{9\zeta^2(3/2)} = 1.2855.
\end{align}
The amplification of charge susceptibility also implies the reduction of magnetic permeability $\mu$ in systems are that particle-vortex duals \cite{Dam-T-Son-2002}. This can be justified by formally identifying $\frac{1}{4\pi}a_{c}$ and $\frac{-1}{4\pi}A_0$ as the source of $B$ and $b_{c}$. In other words, the relationship between chemical potential and the dual magnetic field is analogous to that between $\mb H$ and $\mb B$ in the classical electromagnetism. By writing $H_{\psi} = \frac{a_{c}}{4\pi}$ and $h_{\chi} = -\frac{A_0}{4\pi}$, we can find 
\begin{subequations}
\begin{align*}
\chi_{\chi} &= \frac{\p n_{\chi}}{\p a_{c}} = \frac{1}{(4\pi)^2}\frac{\p B}{\p H_{\psi}}= \frac{1}{(4\pi)^2}\mu_{\psi} = \frac{1}{(4\pi)^2}\left( \frac{2}{3k}\right)^2H_{\psi}\\
 \chi_{\psi} &= \frac{1}{(4\pi)^2}\mu_{\chi} = \frac{1}{(4\pi)^2} (4\pi)^2h_{\chi}.
\end{align*}
\end{subequations}
and as the result of them, 
\begin{align}
\frac{\frac{1}{a_{0}}\chi_{\chi}}{\frac{1}{A_0}\chi_{\psi}} = \frac{\frac{1}{H_{\psi}}\mu_{\psi}}{\frac{1}{h_{\chi}}\mu_{\chi}}.
\end{align}
To further expand the correction to the charge susceptibility, suppose we describe the dual Fermi surface using the phenomenological Fermi liquid theory. This amplification entails that the effective Landau parameter in the scalar channel $F_0$ is negative under gauge field, owing to the relation $\chi=\frac{\chi_{\rm free}}{1 + F_0}$. With
\begin{align}
F_0 = \frac{9\zeta^2(3/2)}{8\pi^2} - 1= -0.2221.
\end{align}
The sign of $F_0$ is nontrivial in that the gauge interaction between charges is repulsive, despite the total Coulomb energy being neutralized. 
\subsection{$N = 2$ duality}
Let us now shift gears to the $N=2$ dualities. Without encountering extra technical difficulties, we can redo the analysis in the previous section to compute the gauge fluctuation-corrected Casimir and magnetic energies. Similar results would follow. In the following, we will mainly focus on the properties specific to Eqs.~\eqref{N=2Duality} and~\eqref{DualN=2}, particularly vacuum expectation value of the fermion condensate in the massless limit highlighted by Eq.~\eqref{condensateEq}.
\subsubsection{$A_0 >0, B=0$}
With this background, $\Psi$ would form a Fermi disk and acquire a finite density. The density $\Psi^{\dag}\Psi$ in turn implies a mean magnetic field $b = \p_1a^2-\p_2a^1$. The other field $\alpha_{\mu}$ detaches from the external probe and does not assume a finite background value.  

Using the Green's function technique, we compute the effective energy (Eq.~\eqref{densityWithMass})
\begin{align}
-& \mathscr E(A_0) = 2\int^{A_0}_0 d\nu\, \mathrm{sgn}(\nu)\int \frac{d^2p}{(2\pi)^2}\theta(\nu^2 - (\mb p^2 + M^2_{\Psi}))\notag\\
= & \frac{\theta(A_0^2-M_{\Psi}^2)}{6\pi}[A_0^3 - 3M_{\Psi}^2A_0 + 2M_{\Psi}^3].
\end{align}
The step function can be dropped because $M_{\Psi}$ would be taken to 0 at the end of operations. Since the $M_{\Psi}$ dependence starts from quadratic order, it is trivially concluded that $\lim_{M_{\Psi}\to 0}\frac{\p\mathscr E}{\p M_{\Psi}} = 0$. The chiral symmetry is not broken for a free 4-component Dirac fermion at finite density.

To investigate $\Gamma$, we employ Eq.~\eqref{N=2AInb}
\begin{align}
\label{N=2AInbMass}  \frac{\p\mathscr E}{\p A_0} & = \frac{b_{c}}{2\pi} =-\frac{1}{2\pi}[A_0^2 - M_{\Psi}^2]<0\notag\\
\Rightarrow  A_0 & = \sqrt{-[b_{c}-M^2_{\Psi}]}.
\end{align}
We have shown $M_{X}\sim -M_{\Psi}$. It suffice to express $\Gamma$ as function of $b_{c}$ and  $M_{\Psi}$ in order to evaluate ${\p\Gamma}/{\p M_{X}}$. Performing the Legendre transform:
\begin{align}\label{N=2GammabcM}
\Gamma[b_{c}, M_{\Psi}] & = -\mathscr E + \frac{b_{c}}{2\pi}A_0 \notag\\
& =  \frac{-1}{3\pi}\sqrt{-(b_{c}-M^2_{\Psi})}^3 + \frac{M^3_{\Psi}}{3\pi}.
\end{align}
When taking the partial derivative, we have to remind ourselves that $b_{c}$ is not entirely $M_{\Psi}$ independent. Physically, $A_0$ is the external knob agnostic of $M_{\Psi}$. Owing to the relation Eq.~\eqref{N=2AInbMass}, $b_{c}$ has to depend on $M_{\Psi}$, and its dependence shall be determined by the condition:
\begin{align}\label{AIndependentM}
0 = \frac{\p}{\p M_{\Psi}}A_0 = \frac{\p}{\p M_{\Psi}}\sqrt{-(b_{c}-M^2_{\Psi})}.
\end{align}
Being mindful of this fact,
\begin{align}
&-\frac{\p\Gamma}{\p M_X}\sim \frac{\p\Gamma}{\p M_{\Psi}} \notag\\
=& \frac{1}{\pi}[-(b_{c}-M_{\Psi}^2)]\frac{\p}{\p M_{\Psi}}\sqrt{-(b_{c}-M^2_{\Psi})} + \frac{1}{\pi}M_{\Psi}^2.
\end{align}
The first term disappears because of~\eqref{AIndependentM}, while the second term vanishes when taking the limit of $M_{\Psi}\to 0$. We conclude $\bar XX$ does not acquire a finite vacuum expectation value in the background $b_{c}$ and in the presence of quantum corrections from $a_{\mu}$ and $\alpha_{\mu}$:
\begin{align}
\lim_{M_X\to 0}\lan \bar XX\ran = 0.
\end{align}
As an alternative perspective, we could express the $\mathscr E[A_0]$ part of the $\Gamma$ fully in terms of $A_0$. By the earlier derivation this term does not produce any finite value after $M_{\Psi}\to 0$. What is left to compute is 
\begin{align}
A_0\lim_{M_{\Psi}\to 0}\frac{\p}{\p M_{\Psi}}\frac{b_{c}}{2\pi}  = 0
\end{align}
by Eq.~\eqref{N=2AInbMass}.
\subsubsection{$B>0, A_0 = 0$}
The $\Psi$ fermions in the magnetic field still form Landau levels and the ground state sits at the charge neutral point. $X$ fermions acquire an average charge density associated with $B$, forming a Fermi disk at the mean-field level, and again subject to fluctuating $a_{\mu}$ and $\alpha_{\mu}$.

$\mathscr E$ is the Euler-Heisenberg Lagrangian with mass. It is typically carried out by dimensional regularization and expressed in terms of Gamma functions. We elaborate on details in the appendix. Utilizing the result Eq.~\eqref{N=2EffL}, it can be formally expanded as:
\begin{align}\label{formalExpansion}
\mathscr E = k_0B^{3/2} + k_1M_{\Psi}B + k_2M^2_{\Psi}B^{1/2}.
\end{align}
This expansion, in terms of the power of $M_{\Psi}/B^{1/2}$, is renormalized such that $M_{\Psi} = 0$ limit corresponds to our previous calculation Eq.~\eqref{magneticB}. Note that, contrary to the case of pure finite density, there is a term in the effective energy linear in $M_{\Psi}$, and therefore \cite{DITTRICH1997182, PhysRevD.52.4718}:
\begin{align}
\lim_{M_{\Psi}\to 0}\bar{\Psi}\Psi = \frac{\p\mathscr E}{\p M_{\Psi}} = k_1B\neq 0.
\end{align}
To solve $B$ with $a_{c}$,
\begin{align}\label{N=2Bina}
\frac{\p\mathscr E}{\p B} = \frac{1}{2\pi}a_{c} = \frac{3}{2}k_0 B^{1/2} + k_1M_{\Psi} + \frac{1}{2}k_2M^2_{\Psi}B^{-1/2}.
\end{align}
The solution in the limit of small $M_{\Psi}$ reads:
\begin{align}
B^{1/2} & = \frac{\frac{a_{c}}{2\pi} - k_1M_{\Psi}+ \sqrt{(\frac{a_{c}}{2\pi} - k_1M_{\Psi})^2 - 3k_0k_2M^2_{\Psi}}}{3k_0}\notag\\
& \approx \frac{2}{3k_0}(\frac{a_{c}}{2\pi} -k_1M_{\Psi}).
\end{align}
Similar to  Eq.~\eqref{AIndependentM}, we have to remind ourselves that $a_{c}$ depends on both $M_{\Psi}$ and $B$ since $B$ is agnostic of $M_{\Psi}$, and the dependence here is inferred by:
\begin{align}\label{BIndependentM}
0 = \frac{\p B^{1/2}}{\p M_{\Psi}} = \frac{\p}{\p M_{\Psi}} \frac{2}{3k_0}(\frac{a_{c}}{2\pi} -k_1M_{\Psi}).
\end{align}
Plugging back the solutions to the Legendre transform, we obtain
\begin{align}
\Gamma = & \frac{4}{27k_0^2}  \left(\frac{a_{c}}{2\pi} -k_1M_{\Psi}\right)^3 \notag\\
& - \frac{2k_2M_{\Psi}^2}{3k_0}\left(\frac{a_{c}}{2\pi} -k_1M_{\Psi}\right).
\end{align}
We observe that there is also a term explicitly linear in $M_{\Psi}$ in the leading cubic term. A naive generalization of the earlier argument would suggest the magnetic catalysis in the free theory induces chiral symmetry breaking by a condensate $\sim  -\frac{4k_1}{9k_0^2}(\frac{a_{c}}{2\pi})^2$ in the U(1)$\times$U(1) QED${}_3$ at finite density. Nevertheless, owing to the dependence required by~\eqref{BIndependentM}, we see
\begin{align}
\lim_{M_X\to 0}\lan\bar XX\ran \sim \frac{\p}{\p M_{\Psi}}\Gamma[a_{c}] = 0.
\end{align}
Let us again examine this from the alternative perspective by writing $\mathscr E$ in terms of $B$ and $M_{\Psi}$. Since $B$ is $M_{\Psi}$ independent, the contribution of this term to the condensate follows our linear term argument and is $-k_1$. The second contribution comes from $\frac{\p a_{c}/(2\pi)}{\p M_{\Psi}}$. By Eq.~\eqref{N=2Bina}
\begin{align*}
 & \lim_{M_X\to 0}\lan\bar{X}X\ran \sim  \lim_{M_{\Psi}\to 0}\frac{\p \Gamma}{\p M_{\Psi}} \notag\\
 =&\lim_{M_{\Psi}\to 0} \left( -k_1 + B\frac{\p}{\p M_{\Psi}}\frac{a_{c}}{2\pi}\right) = -k_1 + k_1 = 0.
\end{align*}
The contribution from $\mathscr E$ is compensated by the coupling of $a_{c}$ to the magnetic field. The exact value of $k_1$ is irrelevant as long as the effective action assumes the form of the $m/B^{1/2}$ expansion~\eqref{formalExpansion}.
\subsubsection{Discussion}
In this section we considered two background field configurations where a free 4-component Dirac field has known solutions, and their particle-vortex duals. The gauge field corrections computed for $N=1$ are directly generalized. Even in the case of finite mass, the interpretation of dual energy still holds. For example, Eqs~\eqref{N=2GammabcM} equals the ground state energy of the Fermi disk up to the Fermi level:
\begin{align*}
2\int\frac{d^2p}{(2\pi)}\theta(A^2_0-(\mb p^2+M_{\Psi}^2))\sqrt{\mb p^2 + M^2_{\Psi}} = \frac{1}{3\pi}[A^3_0 - M^3_{\Psi}].
\end{align*}
Thus, we did not duplicate the derivations. Instead, we focused on the puzzle of spontaneous breakdown of chiral symmetry for $N=2$. In the free theory, on one hand, at finite density without a magnetic field, the chiral symmetry persists. On the other hand, in a constant magnetic field at zero density, the chiral symmetry is broken by the zero modes on the lowest Landau level. The dual theory of the 4-component Dirac field is a U(1)$\times$U(1) QED${}_3$. We derived the effective actions for them and investigated the equations for mass condensation. The conclusion was that neither a constant charge density nor a constant magnetic field is able to create a finite condensate in the massless limit. This might be intuitive for the case of charge density. Remarkably, the magnetic catalysis in the free theory limit is erased by gauge field fluctuation, which, as we reveal, can be attributed to its dual description as a free Fermi disk. We argue that it is a fair comparison even if the free theory does not seem to contain a second U(1) gauge field -- it can be understood as a choice of pure gauge $\mathscr A_{\mu} = 0$ for a coupling $\mathscr A_{\mu}(\bar{\psi}_1\ga^{\mu}\psi_1 - \bar{\psi}_2\bar{\ga}^{\mu}\psi_2)$ in Eq.~\eqref{N=2free}. It naturally implies $\mathscr A_0 = 0$ and $(\p_{\mu}\mathscr A_{\nu}-\p_{\nu}\mathscr A_{\mu})=0$, corresponding to $\alpha_{\mu}$ in the dual theory. 

The reduction of magnetic catalysis in the presence of two U(1) fields might be explained with the following picture. For a free 4-component Dirac field $\Psi = (\psi_1, \psi_2)^T$ of mass $M>0$, in a uniform magnetic field it acquires two Landau levels whose energy values $\pm M$ are independent of the strength of magnetic field. Suppose the ground state is defined by filling up all negative-energy states. The band with energy $-M$ would be filled, whereas the Landau level with energy $M$ is empty. As we lower the mass $M\to 0^+$, two Landau levels become degenerate. The choice of filling up which Landau orbitals leads to spontaneous breakdown of the symmetry and is the origin of magnetic catalysis. Using explicit solutions to the Dirac equation such as Eqs.~\eqref{ansatzM} and~\eqref{ansatzNM}, for these two Landau levels:
\begin{align}
\bar{\Psi}\Psi = \psi^{\dag}_1\sigma_3 \psi_1 - \psi^{\dag}_2\sigma_3\psi_2 \propto -\psi^{\dag}_1\psi_1 + \psi^{\dag}_2\psi_2
\end{align}
because the solution eigenspinors, in the 2-component subspace, have the same $\sigma_3$ eigenvalue. A finite value of $\bar{\Psi}\Psi$ over the zero modes algebraically implies the polarization of charge densities. Switching back to our dual U(1)$\times$U(1) QED${}_3$, an analogous polarization shall be induced if there is magnetic catalysis in the massless limit. Nevertheless, the constraint equation~\eqref{mapping7} forbids any charge density polarization between species, consequently restoring the broken symmetry.
\section{Concluding remark}\label{conclusion}
Exploiting fermionic particle-vortex duality, we developed a machinery to include gauge field fluctuations into the effective action for a class of QED${}_3$ that are dual to free Dirac cones. Using this technique, we showed that the fermionic Casimir energy at finite density and ground energy in a constant magnetic field are both amplified by quantum fluctuations. We also assign physical meanings to each term in the Legendre transform and provide alternative derivations of the results. Moreover, the machinery was also utilized to demonstrate that in the U(1)$\times$U(1) QED${}_3$ with two species of Dirac fermions, there is no spontaneous breakdown of the chiral symmetry caused by the average constant magnetic field seen by both fermion fields, as it is essentially dual to a free Fermi disk of a 4-component Dirac spinor.

We expect this work prototypes a helpful methodology for quantitatively studying (2+1) dimensional QEDs in the infrared with the method of effective action. We focused on a limited number of dualities and background field configurations with either a finite chemical potential or a finite magnetic field in order to simplify the technical processes and highlight some physically interpretable implications for this new technique. Nonetheless, the applicability shall be legitimate for any model that admits a dual free theory of Dirac fermions.

For instance, we can dualize all solvable electromagnetic field configurations, such as turning on an electric field. An intriguing case directly following this work is when $A_0\neq 0$ and $B\neq 0$ simultaneously. The effective action for a free Dirac cone is a sum of $(-1)\times$~\eqref{CasimirA0}, ~\eqref{magneticB}, and the convolution of magnetic field and chemical potential. The latter is oscillatory with frequencies proportional to the filling fraction $2\pi n/B$ \cite{PhysRevD.51.2026, PhysRevD.55.6218} and could possibly be applied to the de Haas-van Alphen effect in the dual theory.

On top of more intricate field configurations, it is fairly straightforward to generalize our examples to a general number of fermion species $N$. Perhaps even more interestingly, we can instead employ fermion-boson dualities to study (2+1) dimensional scalar QEDs which are dual to free Dirac fermions \cite{SEIBERG2016395}. Such QEDs contains integral level Chern-Simons terms for the dynamical fields in order to mutate statistics of the matter fields, suggesting more involved operator correspondences, but the argument leading to the Legendre transform shall still hold given the same structure of the actions.

Despite the fact that not all dual interacting theories have direct correspondences to the most actively studied models or theories, the perspective and the results presented in this work can likely shed some lights on the common features of interacting gauge theories and potential developments of quantitative techniques with (2+1) dimensional field theory dualities.

%\begin{acknowledgments}
%We thank Yu-Ping Lin for comments on the early version of this manuscript.
%\end{acknowledgments}

\appendix
\begin{widetext}
\section{effective action at finite density and no magnetic field}
In this section we compute the functional determinant of a Dirac field at finite density. The method largely follows \cite{PhysRevD.42.2881} except we work in (2+1) dimensions. We consider a 2-component Dirac Lagrangian:
\begin{align}
\mathscr L = \bar{\psi}(i\slashed{\p} +  A_0\ga^0)\psi.
\end{align}
The effective action derived from this Lagrangian is the functional determinant. Raising the determinant to the exponent and subtract the vacuum value, the target to compute amounts to 
\begin{align}\label{LFromDet}
\int d^3x\, \mathscr L_{\rm eff} = -i \left( \mathrm{Tr}\ln[i\slashed{\p} + A_0\ga^0] - \mathrm{Tr}\ln i\slashed{\p}\right),
\end{align}
where $\mathrm{Tr}$ includes spacetime trace $\int d^3x\, \lan x| \cdots | x\ran$ and the gamma matrix trace $\mathrm{tr}$. Using the fact that 
\begin{align}
\mathrm{Tr}\frac{\p}{\p A_0}\ln[i\slashed{\p} + A_0\ga^0] = \int d^3x\, \mathrm{tr}\lan x|[\frac{\ga_0}{i\slashed{\p} + A_0\ga^0}]|x\ran
\end{align}
The effective Lagrangian equals the diagonal element of the inverse operator of $i\slashed{\p} + A_0\ga^0$, or more conventionally the Green's function: 
\begin{align}
(i\slashed{\p} + A_0\ga^0)G(x, y) = \delta^{(3)}(x-y).
\end{align}
At finite density, the solution is given by the following epsilon prescription
\begin{align}
G(x-y;A_0) = \int \frac{d^3p}{(2\pi)^3}\frac{\slashed{\tilde{p}}e^{-ip(x-y)}}{(\tilde{p}_0 + i\epsilon\mathrm{sgn}p_0)^2 - \mb p^2}, \tilde{p} = (p_0 + A_0, \mb p).
\end{align}
Plugging the explicit Green's function into Eq.~\eqref{LFromDet}, it becomes
\begin{align}
\mathscr L_{\rm eff}= -i \int^{A_0}_0 d\nu\, \mathrm{tr}[G(0; \nu)\ga^0] = -2i \int^{A_0}_0 d\nu \frac{d^3p}{(2\pi)^3} \frac{(p_0 + \nu)}{(p_0 + \nu + i\epsilon\mathrm{sgn}(p_0))^2 - \mb p^2}.
\end{align}
The temporal integral can be computed as follows:
\begin{align}
& I = \int^{\infty}_{-\infty} d\omega \frac{\omega + \nu}{(\omega + \nu + i\epsilon\mathrm{sgn}\omega)^2 - \mb p^2} = \int^{\infty}_{-\infty}d\omega \frac{\omega}{(\omega + i\epsilon\mathrm{sgn}(\omega - \nu))^2 - \mb p^2}=  - \int^{\infty}_{-\infty}d\omega \frac{\omega}{(\omega + i\epsilon\mathrm{sgn}(\omega + \nu))^2 - \mb p^2}\notag\\
= & \frac{1}{2}\int^{\infty}_{-\infty}d\omega \omega \frac{[\omega + i\epsilon\mathrm{sgn}(\omega + \nu)]^2 - [\omega + i\epsilon\mathrm{sgn}(\omega - \nu)]^2}{[(\omega + i\epsilon\mathrm{sgn}(\omega - \nu))^2 - \mb p^2][(\omega + i\epsilon\mathrm{sgn}(\omega + \nu))^2 - \mb p^2]}\notag\\
= & \frac{1}{2}\int^{\infty}_{-\infty}d\omega \omega^2 \frac{2i\epsilon[\mathrm{sgn}(\omega + \nu)-\mathrm{sgn}(\omega- \nu)]}{[(\omega + i\epsilon\mathrm{sgn}(\omega - \nu))^2 - \mb p^2][(\omega + i\epsilon\mathrm{sgn}(\omega + \nu))^2 - \mb p^2]}\notag\\
= & \int^{\infty}_{-\infty}d\omega\omega^2 \frac{2i\epsilon\mathrm{sgn}(\nu)\theta(|\nu|-\omega)\theta(|\nu| + \omega)}{[(\omega +i\epsilon)^2-\mb p^2][(\omega - i\epsilon)^2-\mb p^2]} =  \int^{|\nu|}_{-|\nu|} d\omega \frac{2i\epsilon\omega^2\mathrm{sgn}(\nu)}{[(\omega + i\epsilon)^2-\mb p^2][(\omega - i\epsilon)^2-\mb p^2]} \notag\\
= & \int^{|\nu|}_{0} d\omega \frac{4i\epsilon\omega^2\mathrm{sgn}(\nu)}{[(\omega + i\epsilon)^2-\mb p^2][(\omega - i\epsilon)^2-\mb p^2]} =  \int^{|\nu|}_0 d\omega \omega\, \mathrm{sgn}(\nu)\left[ \frac{1}{(\omega - i\epsilon)^2-\mb p^2}- \frac{1}{(\omega + i\epsilon)^2-\mb p^2}\right] \notag\\
= & \frac{1}{2}\mathrm{sgn}(\nu)\int^{|\nu|^2}_0 d\omega^2 \left[ \frac{1}{\omega^2 - \mb p^2 - i\epsilon} - \frac{1}{\omega^2 - \mb p^2 + i\epsilon}\right] = i\pi \mathrm{sgn}(\nu)\theta(\nu^2 - \mb p^2).
\end{align}
Therefore, 
\begin{align}
\mathscr L_{\rm eff}(\nu) = \int^{A_0}_0 d\nu\, \mathrm{sgn}(\nu)\int \frac{d^2p}{(2\pi)^2}\theta(\nu^2 - \mb p^2) = \int^{A_0}_0d\nu\, \mathrm{sgn}(\nu)\frac{1}{4\pi}\nu^2 = \frac{1}{12\pi}\mathrm{sgn}(A_0)A_0^3 = \frac{1}{12\pi}|A_0|^3.
\end{align}
To generalize the effective Lagrangian to a 4-component massive Dirac field, we simply have to replace $\mb p^2\to \mb p^2 + m^2$ in the above and incorporate the multiplicity of $2$.
\begin{align}
\label{densityWithMass}& \mathscr L_{\rm eff}(A_0) = 2 \int^{A_0}_0 d\nu\, \mathrm{sgn}(\nu)\int \frac{d^2p}{(2\pi)^2}\theta(\nu^2 - (\mb p^2 + m^2))\notag\\
= &2 \int^{A_0}_0 d\nu\, \mathrm{sgn}(\nu)\theta(\nu^2 - m^2)\int^{\sqrt{\nu^2-m^2}}_0 \frac{d^2p}{(2\pi)^2} = 2\int^{A_0}_0 d\nu\, \mathrm{sgn}(\nu)\frac{\nu^2-m^2}{4\pi}\theta(\nu^2-m^2)\notag\\
= & \frac{\theta(A_0^2-m^2)}{2\pi}\int^{|A_0|}_md\nu (\nu^2-m^2) = \frac{\theta(A_0^2-m^2)}{2\pi}\left(\frac{\nu^3}{3}-m^2\nu\right)\bigg|^{|A_0|}_m\notag\\
= & \frac{\theta(A_0^2-m^2)}{6\pi}[|A_0|^3 - 3m^2|A_0| + 2m^3] = \frac{\theta(A_0^2-m^2)|A_0|^3}{6\pi}\left( 1 - 3\frac{m^2}{|A_0|^2} + \frac{2m^3}{|A_0|^3}\right)
\end{align}
\section{effective action in a constant magnetic field at zero density}
\subsection{Setting up proper time representation}
In this section we employ the Schwinger proper time method to compute the Euler-Heisenberg Lagrangian in (2+1) dimensions. To start, we rewrite the logarithm of the Dirac operator in the proper time representation. Using the trick,
\begin{align}
 \mathscr L_{\rm eff} = -i \mathrm{Tr}\ln[i\slashed{\p} + \slashed A + \eta]\Rightarrow \frac{\p\Gamma_{\rm eff}}{\p \eta} = -i \mathrm{Tr}[\frac{1}{i\slashed{\p} + \slashed A + \eta}] =  -\frac{1}{2}i \frac{\p}{\p\eta} \mathrm{Tr}\left[ \ln[-\slashed D^2 - \eta]\right]
\end{align}
and the identity $\ln(A+i\epsilon) =  -\int^{\infty}_{s_0} \frac{ds}{s}e^{i(A+i\epsilon)}$,
\begin{align}
\mathscr L_{\rm eff} = -\frac{i}{2}\left( \mathrm{Tr}\ln[-\slashed D^2] - \mathrm{Tr}\ln[-\slashed{\p}^2]\right) = \frac{i}{2}\left[ \int^{\infty}_0 \frac{ds}{s}\mathrm{Tr}[e^{-i\slashed{D}^2} ]- \int^{\infty}_0 \frac{ds}{s}\mathrm{Tr}[e^{-i\slashed{\p}^2}]\right]
\end{align}
up to an additive constant. Expanding the exponent, we can identify the Hamiltonian
\begin{align}
 \slashed{D}^2 = \frac{1}{2}g^{\mu\nu}\{ \p_{\mu} - iA_{\mu}, \p_{\nu} - iA_{\nu}\} - \frac{i}{4}[\ga^{\mu}, \ga^{\nu}]F_{\mu\nu} = -\pi_{\mu}\pi^{\mu} - \frac{1}{2}\sigma_{\mu\nu}F^{\mu\nu}.
\end{align}
This decomposition allows us to pull the constant electromagnetic field out of the spacetime trace:
\begin{align}
\frac{i}{2}\int^{\infty}_0 \frac{ds}{s}\mathrm{Tr}[e^{-i\slashed{D}^2s}] = \frac{i}{2}\int d^3x\, \int^{\infty}_0\frac{ds}{s}e^{-im^2s}\mathrm{tr}[e^{\frac{i}{2}\sigma_{\mu\nu}F^{\mu\nu}s}]\lan x| e^{-i(-\pi_{\mu}\pi^{\mu})s}|x\ran,
\end{align}
where we insert a small mass $m^2$ to indicate the contour orientation. The main task in the integrand is the diagonal matrix element \footnote{We use $y$ to denote the eigenvalue of position operator and preserve $x$ for the operators.}
\begin{align}
\lim_{y\to y'}\lan y | e^{-i(-\pi^{\mu}\pi_{\mu})s}|y'\ran.
\end{align}
\subsection{Solving equation of motion}
Let us write
\begin{align}
\lan y|e^{-i(-\pi_{\mu}\pi^{\mu})s}|y'\ran =\lan y;0|e^{-i(-\pi_{\mu}\pi^{\mu})s}|y'; 0\ran := \lan y; 0|y';s\ran.
\end{align}
It obeys the differential equation
\begin{align}
i\p_s\lan y;0|y';s\ran = \lan y;0|e^{-i(-\pi_{\mu}\pi^{\mu})s}(-\pi^{\mu}\pi_{\mu})|y'; 0\ran 
\end{align}
As long as we can solve $\pi^{\mu}$ in terms of the position operator $x^{\mu}$, using 
\begin{align}
& x^{\mu}(0)|y'; 0\ran = {y'}^{\mu}|y'; 0\ran\notag\\
& \lan y; 0|e^{-i(-\pi_{\nu}\pi^{\nu})s}x^{\mu}(s) = \lan y; 0|x^{\mu}(0)e^{-i(-\pi^{\nu}\pi_{\nu})s} = y^{\mu} \lan y; 0|e^{-i(-\pi^{\nu}\pi_{\nu})s},
\end{align}
the amplitude on the right-hand side can be evaluated. Exploiting the canonical commutation relations $[x^{\mu}, \pi^{\nu}] = -ig^{\mu\nu}$ and $ [\pi^{\mu}, \pi^{\nu}] = i F^{\mu\nu}$. The equations of motion are given by
\begin{align}
& \frac{dx^{\mu}}{ds} = 2\pi^{\mu}\\
& \frac{d\pi^{\mu}}{ds} = -2F^{\mu}\!_{\nu}\pi^{\nu}.
\end{align}
Thus,
\begin{align}
\pi^{\mu}(s) = [e^{-2Fs}]^{\mu}\!_{\nu}\pi^{\nu}(0).
\end{align}
It is worth emphasizing that, in the rest of the section, whenever we omit the indices for symbolic simplicity, we mean $\pi \to \pi^{\mu}, x\to x^{\mu}, F\to F^{\mu}\!_{\nu}$. By $F^{\mu}\!_{\nu} =  -F_{\nu}\!^{\mu}$,
\begin{align}
\frac{dx^{\mu}}{ds} = 2\pi^{\mu} = 2(e^{-2Fs})^{\mu}\!_{\nu}\pi^{\nu}(0)\Rightarrow  x(s)-x(0) = 2F^{-1}e^{-Fs}\sinh Fs\pi(0).
\end{align}
The momenta operators can thus be expressed using the position operators
\begin{align}
& \pi(0) = \frac{Fe^{Fs}}{2\sinh Fs}(x(s)-x(0))\\
& \pi(s) = e^{-2Fs}\pi(0) = \frac{Fe^{-Fs}}{2\sinh Fs}(x(s)-x(0)),
\end{align}
and the Hamiltonian is similarly rephrased as
\begin{align}
& -\pi_{\mu}(s)\pi^{\mu}(s) = - (x(s)-x(0))^{\mu}\left( \frac{F^2}{4\sinh^2Fs}\right)_{\mu\nu}(x(s)-x(0))^{\nu}\notag\\
:= & - [x^{\mu}(s)K_{\mu\nu}x^{\nu}(s) - 2x^{\mu}(s)K_{\mu\nu}x^{\nu}(0) + x^{\mu}(0)K_{\mu\nu}x^{\nu}(0) - K_{\mu\nu}[x^{\mu}(0), x^{\nu}(s)]].
\end{align}
The quantum mechanical property is encoded in the commutator 
\begin{align}
&-K_{\mu\nu}[x^{\mu}(0), x^{\nu}(s)]=i \left( \frac{F^2}{4\sinh^2 Fs}\frac{2e^{-Fs}\sinh Fs}{F}\right)_{\mu}\!^{\mu} = i\left( \frac{Fe^{-Fs}}{2\sinh Fs}\right)_{\mu}\!^{\mu} = i \left( \frac{F}{2}\coth Fs\right)_{\mu}\!^{\mu}
\end{align}
using ${e^{-Fs}}/{\sinh Fs} = \coth Fs - 1.$ As the result,
\begin{align}
-\pi^{\mu}\pi_{\mu} = -[x^{\mu}(s)K_{\mu\nu}x^{\nu}(s) - 2x^{\mu}(s)K_{\mu\nu}x^{\nu}(0) + x^{\mu}(0)K_{\mu\nu}x^{\nu} + \frac{i}{2}\mathrm{tr}_F(F\coth Fs)]
\end{align}
and hence 
\begin{align}
&- [(y-y')^{\mu}K_{\mu\nu}(y-y')^{\nu} + \frac{i}{2}\mathrm{tr}_FF\coth Fs]\lan y; 0|y'; s\ran = i \p_s\lan y; 0|y'; s\ran\\
\Rightarrow & \lan y; 0| y'; s\ran = C(y, y')\exp\left( -i (y-y')\frac{F\coth Fs}{4}(y-y') - \frac{1}{2}\mathrm{tr}_F\ln \frac{\sinh Fs}{F}\right)\notag\\
= & \frac{C(y, y')}{s^{3/2}}\exp\left( -i (y-y')\frac{F\coth Fs}{4}(y-y') - \frac{1}{2}\mathrm{tr}_F\ln \frac{\sinh Fs}{Fs}\right).
\end{align}
The notation $\mathrm{tr}_F$ refers to summing over $\mu, \nu$ indices for the electromagnetic field tensor, and it is one of the main sources of spacetime dimensions dependence in this calculation as we can see from the coefficient of $\ln s$. We would also like to point out a subtlety that when integrating $\coth Fs$ the $F^{-1}$ in the final logarithm is an integral constant. This is separated from the rest of the constant parts in order to facilitate $F\to 0$ limit more easily. To determine the coefficient $C(y, y')$, we impose boundary conditions to match physical quantities. Recall that from the basic quantum mechanics, $ \lan x^{\mu}|p_{\mu}|\psi\ran = i \p_{\mu}\lan x^{\mu}|\psi\ran.$
\begin{align}
& \left[ -i\frac{\p}{\p{y'}^{\mu}} + A_{\mu}(y')\right]\lan y; 0|y'; s\ran = \lan y; 0|e^{-i(-\pi^{\nu}\pi_{\nu})s}\pi_{\mu}(0)|y'; 0\ran\\
& \left[ i\frac{\p}{\p{y}^{\mu}} + A_{\mu}(y)\right]\lan y; 0|y'; s\ran = \lan y; 0|\pi_{\mu}(0)e^{-i(-\pi^{\nu}\pi_{\nu})s}|y'; 0\ran = \lan y; 0|e^{-i(-\pi^{\nu}\pi_{\nu})s}\pi_{\mu}(s)|y'; 0\ran.
\end{align}
Plugging the semi-solution into the above boundary conditions,
\begin{align}
 -i \frac{\p}{\p y'_{\mu}}C(y,y') & = [-A^{\mu}(y') + \frac{1}{2}F^{\mu}\!_{\nu}(y-y')^{\nu}]C(y,y')\\
 i \frac{\p}{\p y_{\mu}}C(y,y') & = [-A^{\mu}(y) - \frac{1}{2}F^{\mu}\!_{\nu}(y-y')^{\nu}]C(y,y').
\end{align}
These two equations have the general solution
\begin{align}
\ln C(y, y') = i \int^{y}_{y'} dz_{\lambda}[A^{\lambda}(z) + \frac{1}{2}F^{\lambda}\!_{\nu}(z-y')^{\nu}].
\end{align}
By choosing $y\to y'$ to be a straight line,
\begin{align}
\lan y; 0|y'; s\ran = \frac{C}{s^{3/2}}e^{i \int^y_{y'}dz A}\exp\left( - i(y-y')_{\mu}\left( \frac{F\coth Fs}{4}\right)^{\mu}\!_{\nu}(y-y')^{\nu} - \frac{1}{2}\mathrm{tr}_F\ln\frac{\sinh Fs}{Fs}\right).
\end{align}
To determine $C$, we demand that in the limit of $F\to 0$ this amplitude should reduce to the free particle propagator. First we turn off the electromagnetic field 
\begin{align}
\lan y; 0|y'; s\ran \to \frac{C}{s^{3/2}}\exp\left( -i (y-y')_{\mu}\frac{1}{4s}(y-y')^{\mu}\right) = \frac{C}{s^{3/2}}\exp\left[ -i\frac{(t-t')^2}{4s} + i \frac{(\mb y-\mb y')^2}{4s}\right].
\end{align}
Imposing the normalization condition implies 
\begin{align}
1 = \int d^3\Delta y \frac{C}{s^{3/2}} \exp\left[ -i\frac{(t-t')^2}{4s} + i \frac{(\mb y-\mb y')^2}{4s}\right] \Rightarrow C = \frac{e^{-i\pi/4}}{(4\pi)^{3/2}}.
\end{align}
Here we have yet another subtlety depending on the metric signature. If this calculation is performed in the Euclidean signature, the exponents would all have the same sign and the final $C$ would differ by a phase $e^{i\pi}= -1$. The final form of the transition amplitude reads 
\begin{align}
\lan y; 0|y';s\ran = \frac{e^{-i\pi/4}}{(4\pi s)^{3/2}}e^{i\int_{y'}^y dz^{\mu}A_{\mu}}\exp\left[ -i (y-y')_{\mu}\left( \frac{F\coth Fs}{4}\right)^{\mu}\!_{\nu}(y-y')^{\nu}- \frac{1}{2}\mathrm{tr}_F\ln\frac{\sinh Fs}{Fs}\right].
\end{align}
Letting $y\to y'$,
\begin{align}
\lan y| e^{-i(-\pi_{\mu}\pi^{\mu})s}|y\ran = \frac{e^{-i\pi/4}}{(4\pi s)^{3/2}}\exp\left( -\frac{1}{2}\mathrm{tr}_F\ln\frac{\sinh Fs}{Fs}\right)
\end{align}
and consequently the effective action
\begin{align}
\mathscr L_{\rm eff} = \frac{ie^{-i\pi/4}}{16\pi^{3/2}}\int^{\infty}_0 \frac{ds}{s^{5/2}}\mathrm{tr}e^{\frac{i}{2}\sigma_{\mu\nu}F^{\mu\nu}s}e^{-im^2s}\exp\left( -\frac{1}{2}\mathrm{tr}_F\ln\frac{\sinh Fs}{Fs}\right).
\end{align}
We can eventually evaluate this for the configuration $F_{12} = -B$ and $\sigma_{12} = \sigma_z$. The gamma matrix and spacetime index traces can be easily evaluated by diagonalize a 2 by 2 and a 3 by 3 matrix respectively. After subtracting the vacuum value in the limit of $B=0$,
\begin{align}\label{N=1LeffIntegral}
\mathscr L_{\rm eff} = \frac{e^{i\pi/4}}{8\pi^{3/2}}\int^{\infty}_0 \frac{ds}{s^{5/2}}e^{-im^2s}(Bs\cot Bs - 1).
\end{align}

\subsection{Computing proper time integral in the massless limit}
Following the result in the above, given the sign of $m^2$, we have to close the contour in the lower-right quadrant and come back from the negative imaginary axis. Since the integral vanishes in the infinity, a zero sum round trip implies 
\begin{align}
\frac{e^{i\pi/4}}{8\pi^{3/2}}\int^{\infty}_0 \frac{ds}{s^{5/2}}e^{-im^2s}(|B|s\cot|B|s - 1) = \frac{e^{i\pi/4}}{8\pi^{3/2}}\int^{-i\infty}_0 \frac{ds}{s^{5/2}}e^{-im^2s}(|B|s\cot|B|s - 1).
\end{align}
Let $s = -iy$. It becomes
\begin{align}
\frac{-1}{8\pi^{3/2}}\int^{\infty}_0 \frac{dy}{y^{5/2}}e^{-m^2y}(|B|y\coth|B|y- 1).
\end{align}
Most literature goes straight to the substitution of variable and therefore might be unclear. Using the series expansion 
\begin{align}
\coth z = \sum_{k = -\infty}^{\infty}\frac{z}{\pi^2k^2 + z^2},
\end{align}
and take the limit $m\to 0^+$, the above becomes  
\begin{align}
 - \frac{|B|^{3/2}}{4\pi^{3/2}}\int^{\infty}_0 \frac{ds}{s^{5/2}} \sum_{n = 1}^{\infty}\frac{s^2}{s^2 + n^2\pi^2}.
\end{align}
Each summand can be integrated easily 
\begin{align}
\int^{\infty}_0 \frac{ds}{s^{1/2}}\frac{1}{s^2 + n^2\pi^2} = \int^{\infty}_0 2 dx\frac{1}{x^4 + n^2\pi^2} = \int^{\infty}_{-\infty}dx\, \frac{1}{x^4 + n^2\pi^2} =\frac{1}{\sqrt{2\pi}n^{3/2}}.
\end{align}
As a result,
\begin{align}
\mathscr L_{\rm eff} = \frac{-|B|^{3/2}}{4\pi^{3/2}}\frac{1}{\sqrt{2\pi}}\sum_{n = 1}\frac{1}{n^{3/2}} = -\frac{|B|^{3/2}\zeta(3/2)}{4\sqrt{2}\pi^2}.
\end{align}

\subsection{Effective action in constant magnetic field with finite mass for $N=2$ theory at zero density}
We assume $B, m>0$ without loss of generality. The effective action for $N=2$ is given by $2\times$~\eqref{N=1LeffIntegral}:
\begin{align}
& \mathscr L_{\rm eff} = -\frac{1}{4\pi^{3/2}}\int^{\infty}_0 \frac{ds}{s^{5/2}}(sB\coth Bs - 1)e^{-m^2s},\ s = B^{-1}t,\notag\\
= & - \frac{B^{3/2}}{4\pi^{3/2}}\int^{\infty}_0 dt\, e^{-\frac{m^2}{B}t}\left( \frac{1}{t^{3/2}}\coth t - \frac{1}{t^{5/2}}\right)\notag\\
= & -\frac{B^{3/2}}{4\pi^{3/2}}\left[ \int^{\infty}_0 dt\, \frac{e^{-\frac{m^2}{B}t}}{t}t^{-1/2}\coth t - \int^{\infty}_0 dt\, \frac{e^{-\frac{m^2}{B}t}}{t}t^{-3/2}\right]\notag\\
= & -\frac{B^{3/2}}{4\pi^{3/2}}\left[ \left( 2^{3/2}\zeta(-1/2, m^2/(2B)) - \sqrt{\frac{m^2}{B}}\right)\Gamma(-1/2) - (m^2/B)^{3/2}\Gamma(-3/2)\right]\notag\\
= & -\frac{1}{4\pi^{3/2}}\left[ 2^{3/2}\zeta(-1/2, \frac{m^2}{2B})B^{3/2}- Bm\right]\Gamma(-1/2)  + \frac{m^3}{4\pi^{3/2}}\Gamma(-3/2).
\end{align}
The integral identities \cite{Gradshteyn:1702455}
\begin{subequations}
\begin{align}
& \int^{\infty}_0 dx\, x^{\mu-1}e^{-\beta x} = \frac{1}{\beta^{\mu}}\Gamma(\mu)\\
& \int^{\infty}_0dx\, x^{\mu -1}e^{-\beta x}\coth x = \Gamma(\mu)\left[ 2^{1-\mu}\zeta(\mu, \beta/2) - \beta^{-\mu}\right]
\end{align}
\end{subequations}
were quoted to evaluate the integrals. We note that the Gamma and zeta functions serve regularization purpose. We should technically insert a cutoff near $s = 0$. 
Formally in the limit $m^2\ll B$, we can approximate
\begin{align}
\label{N=2EffL}
\mathscr L_{\rm eff}\approx -\frac{1}{4\pi^{3/2}} \left( 2^{3/2}\zeta(-1/2, 0)B^{3/2} + 2^{1/2}\zeta'(-1/2, 0)m^2B^{1/2}- Bm\right)\Gamma(-1/2) + \frac{m^3}{4\pi^{3/2}}\Gamma(-3/2).
\end{align}
The condensation in the massless limit can be evaluated 
\begin{align}
\lan\bar{\Psi}\Psi\ran = -\lim_{m\to 0}\frac{\p\mathscr L_{\rm eff}}{\p m} = -\frac{B}{4\pi^{3/2}}\Gamma(-1/2) = \frac{B}{2\pi}.
\end{align}
\end{widetext}
\section{Landau level solution}
In this section we present an explicit representation of the lowest Landau level wave functions for a free 4-component Dirac field $\Psi$ in order to support our discussion in the main text. We adopt the following reducible representation for the gamma matrices:
\begin{subequations}
\begin{align}
& \Gamma^0 = \begin{pmatrix} \sigma_3 & 0 \\ 0 & -\sigma_3\end{pmatrix} = \begin{pmatrix} 1 & 0 & 0 & 0 \\ 0 & -1 & 0 & 0 \\ 0 & 0 & -1 & 0 \\ 0 & 0 & 0 & 1\end{pmatrix}\\
& \Gamma^1 = \begin{pmatrix} i \sigma_1 & 0 \\ 0 & -i\sigma_1\end{pmatrix} = \begin{pmatrix} 0 & i & 0 & 0 \\ i & 0 & 0 & 0 \\ 0 & 0 & 0 & -i \\ 0 & 0 & -i & 0 \end{pmatrix}\\
& \Gamma^2 = \begin{pmatrix} i \sigma_2 & 0 \\ 0 & -i\sigma_2 \end{pmatrix} = \begin{pmatrix} 0 & 1 & 0 & 0 \\ -1 & 0 & 0 & 0\\ 0 & 0 & 0 & -1\\ 0 & 0 & 1 & 0\end{pmatrix}.
\end{align}
\end{subequations}
The gauge field configuration is given by the Landau gauge $A_{\mu} = \delta_{\mu1}By = \delta_{\mu 1}Bx^2$. The massive Dirac equation reads
\begin{align}
 & [i\Gamma^{\mu}[\p_{\mu}- i A_{\mu}] - M]\Psi  \notag\\
 = & [i\Gamma^0\p_0 +i\Gamma^1\p_1 + \Gamma^1By + i\Gamma^2\p_2 - M]\Psi = 0.
\end{align}
The solutions for general Landau levels can be solved by taking the square of the Dirac operator \cite{PhysRevD.52.4718}. For our purpose, it suffices to verify our ansatzs satisfy the desired eigenvalue equations. Introducing the magnetic length $\ell = B^{-1/2}$ and the wavenumber $k = 2\pi n/L_x$ along the $x$-direction, the following two wave functions 
\begin{subequations}
\begin{align}
\label{ansatzM}& u_{M} = e^{-iM t}e^{ikx}e^{-\frac{1}{2}(y/\ell - k\ell)^2}\begin{pmatrix} 0 \\ 0 \\ 0 \\ 1\end{pmatrix}\\
\label{ansatzNM}& u_{-M} = e^{iM t}e^{ikx}e^{-\frac{1}{2}(y/\ell - k\ell)^2}\begin{pmatrix} 0 \\ 1 \\ 0 \\ 0\end{pmatrix},
\end{align}
\end{subequations}
that satisfy $[i\Gamma^{\mu}[\p_{\mu}- i A_{\mu}] - M]u_{\pm M} = 0.$ We can see explicitly in each 2-component subspace only the upper or lower entry has value, and hence $\bar{\psi}\psi$ and $\psi^{\dag}\psi$ are proportional using these wave functions.

\bibliography{citation}
\end{document}